\documentclass[twocolumn,prl,superscriptaddress,reprint]{revtex4-1}

\usepackage{palatino}
\usepackage[latin1]{inputenc}
\usepackage{epsf}
\usepackage{amsmath,amssymb}
\usepackage{latexsym}
\usepackage{calc}
\usepackage{color}
\usepackage{shadow}
\usepackage{epsfig}

\def\bea{\begin{eqnarray}}
\def\eea{\end{eqnarray}}
\def\ben{\begin{equation}}
\def\een{\end{equation}}
\def\benu{\begin{enumerate}}
\def\enu{\end{enumerate}}

\def\mT{\mathcal{T}}

\def\sss{\scriptscriptstyle\rm}

\def\1var{(\bx_1...\bx\N)}

\def\br{{\bf r}}

\def\bx{{\br t}}

\def\s{_{\sss S}}
\def\xc{_{\sss XC}}
\def\Hx{_{\sss HX}}

\def\Hxc{_{\sss HXC}}

\def\N{_{\sss N}}
\def\H{_{\sss H}}

\def\ext{_{\rm ext}}

\begin{document}
\title{Time-resolved spectroscopy in time dependent density functional theory: An exact condition}
\author{Johanna I. Fuks, Kai Luo, Ernesto D. Sandoval, and Neepa T. Maitra}
\affiliation{Department of Physics and Astronomy, Hunter College and the Graduate Center of the City University of New York, 695 Park Avenue, New York, New York 10065, USA}

\date{\today}
\pacs{}

\begin{abstract}
A fundamental property of a quantum system driven by an external
field is that when the field is turned off the positions of its
response frequencies are independent of the time at which the field is
turned off. We show that this leads to an exact condition for the
exchange-correlation potential of time-dependent density functional
theory. The Kohn-Sham potential typically continues to evolve after the field is
turned off, which leads to time-dependence in the response frequencies of the Kohn-Sham response function. 
The exchange-correlation kernel must cancel out this time-dependence.  The
condition is typically violated by approximations currently in use, as we
demonstrate by several examples, which has severe consequences for
their predictions of time-resolved spectroscopy.

\end{abstract}
\maketitle 
Time-resolved spectroscopies are increasingly being used to characterize and
analyze processes in molecules and solids. Applying an ultrafast pump pulse to create a non-stationary state, which is then monitored in time by a probe pulse is a central technique in the field of femtochemistry~\cite{Zewail00}, and has revolutionized our  understanding of chemical reactions and photo-induced processes in a wide range of systems including biological molecules and nanoscale devices. 
%Pure Appl. Chem., Vol. 72, No. 12, pp. 2219–2231, 2000, Zewail, ``Femtochemistry. Past, present, and future''
Until recently, experiments primarily probed ionic
dynamics 
%and operated largely in the Born-Oppenheimer regime - not sure.
 where time-resolved spectra reflect changes in the ionic
 configuration during a reaction~\cite{PM92}. The recent advent of attosecond
 pulses enables pump-probe experiments at the time-scale of electron
 dynamics~\cite{KI09}, allowing investigations of processes on the electronic
 time-scale, and revealing a wealth of new phenomena and
 new possibilities for characterizing a system.
%Note ultrafast, ultrashort refers to ps pulse and used in femtochem..

A scalable theoretical method to model electron dynamics, reliable beyond the perturbative regime, is crucial to  simulate and interpret experimental results, and to suggest new experiments and materials to study. 
%Furthermore the recent ultrashort ultrafast pulses offer attosecond (1 as$=10^-8$s) time-resolution,  at this timescale the molecular structural changes are eventually frozen ({\it cite: Progress in Ultrafast Intense Laser Science XI
% By Kaoru Yamanouchi, Chang Hee Nam and Philippe Martin}), 
%allowing investigations of pure electron dynamics,
%revealing a wealth of new phenomena~\cite{KI09} and new possibilities for characterizing a system.
Time-dependent density functional theory (TDDFT), an exact reformulation of
many-electron quantum mechanics, stands out with its balance between accuracy and computational
cost~\cite{RG84,TDDFTbook12,Carstenbook}. 
Non-interacting
electrons evolve in a one-body potential such that the exact one-body
density $n(\br,t)$ of the true system is reproduced.
%; the dipole and other moments of the density response are therefore reproduced to all orders.
However in practice the exchange-correlation (xc) contribution to the
potential must be approximated as a functional of the density  and the
initial interacting and Kohn-Sham (KS) states, $\Psi_0,\Phi_0$ respectively: $v\xc[n;\Psi_0,\Phi_0](t)$. Most TDDFT calculations
nowadays use adiabatic functionals, which depend exclusively on the
instantaneous density, input into a ground-state functional:
$v\xc^A[n;\Psi_0,\Phi_0](t) = v\xc^{\rm g.s.}[n(t)]$. 

TDDFT has been extensively and successfully applied to model the linear response of large systems and to elucidate experiments in the non-perturbative regime, 
%in physics and chemistry but has been useful in the non-perturbative regime to interpret experiments, 
e.g. Refs.~\cite{SSYI12,Carlo13,FB11,JBHMDRC13},
including coherent phonon generation, strong-field and thermal ionization,
harmonic generation, and exploring photovoltaic materials, to name a
few.  At the same time, however, recent work on small systems where numerically-exact or high-level wavefunction
methods are
applicable, has shown
that the approximate TDDFT functionals can yield significant errors in
their predictions of the dynamics~\cite{EFRM12,LFSEM14,TGK08,RG12,RB09,FERM13,FHTR11,RN11,HTPI14,RN12,RN12c}, and sometimes they
fail even qualitatively~\cite{FERM13,FHTR11,RB09,RN11,RN12,RN12c,HTPI14}.  
%Hybrid functionals mix in a fraction of exact exchange, which, via
%orbital-dependence, include some non-local density-dependence, but not
%the right type to cure the problems in the cases studied.

A critical aspect of non-equilibrium dynamics are the response
frequencies of the system, since these play a crucial role in the
 response to an applied field, and in interferences in the
dynamics. As pointed out recently, approximate functionals  yield erroneous time-dependent electronic structure when subject to external fields~\cite{RN12,RN12c,GBCWR13}. This spurious ``peak shifting'' makes TDDFT simulations of resonant coherent control very challenging~\cite{RN12}, and the interpretation of time-resolved spectroscopic simulations difficult.

 Let a ``pumped system'' refer to a system 
which has been driven out of its ground-state by an external field
for time $\mT$, after which the field is turned off. 
In this paper we derive an exact condition
that the xc functional must satisfy in order to respect a fundamental property of the response frequencies of the pumped system: For times $\mT$ short enough that ionic motion can be neglected, its response frequencies
 are independent of $\mT$. The
oscillator strengths may change in strength and sign but the response
frequencies remain constant.  
We define the response frequencies via poles
in the density-density linear response function evaluated about an
arbitrary state in the absence of any externally applied fields.  
Most
TDDFT functionals currently in use violate this condition, with severe
implications for the modeling of time-resolved spectroscopy.  Several
model examples are given to illustrate the impact of the violation on
dynamics, including examples of an adiabatic functional that, despite inaccurate response frequencies, approximately satisfies this condition, and consequently yields accurate dynamics.

%Consider a pumped system where the nuclei may be treated as stationary, e.g.  time-scales short enough that nuclear motion can be neglected. 
After the field is turned off at time $\mT$, then, treating the nuclei as stationary, the Hamiltonian is static, 
$\hat{H}^{(0)} = \hat{T} + \hat{W} + \hat{v}\ext^{(0)}$, the sum of the kinetic energy, electron-interaction, and
electron-nuclear interaction operators respectively. 
The electronic state can be expanded in terms of the
eigenstates $\Psi_n$: $\Psi(t\ge\mT) = \sum_n c_n(\mT)\Psi_n
e^{-iE_n(t-\mT)}$, with $\hat{H}^{(0)} \Psi_n= E_n \Psi_n$.
We denote the density of this state  as $n^{(0)}_\mT(\br,t)$, defined for times $t\ge\mT$. Throughout the paper, the superscript ${(0)}$ indicates a quantity in the absence of external fields. 
We define a non-equilibrium response function to describe the
density response to a perturbation $\delta v\ext(\br,t)$ (probe) applied after time $\mT$:
\ben
\tilde\chi[n^{(0)}_\mT; \Psi(\mT)](\br,\br',t,t') = \left.\frac{\delta n(\br,t)}{\delta v\ext(\br',t')}\right\vert_{n^{(0)}_\mT,\Psi(\mT)}\;\;.
\label{eq:chidef}
\een
Here the functional dependences on the left-hand-side follow from the Runge-Gross theorem~\cite{RG84}, 
considering the onset of the free-evolution ($t=\mT$) as the initial time. (Note that Eq.(\ref{eq:chidef}) applies for the response of any arbitrary state $\Psi(\mT)$, not just those reached by a pump field).
%A Lehmann expansion for the density-density response function $\tilde\chi[n^{(0)}_\mT, \Psi(\mT)](\br,\br',t,t')$ can be obtained, 
%by expanding the wavefunction at time $\mT$ in terms of the interacting eigenstates, $\Psi_n$: $\Psi(\mT) = \sum_n c_n(\mT)\Psi_n$. 
Following derivations in standard linear response theory~\cite{gvbook2005} but generalized to an arbitrary initial state,  $\tilde\chi[n^{(0)}_\mT; \Psi(\mT)](\br,\br',t,t') = -i\theta(t-t')\langle \Psi(\mT) \vert [\hat{n}(\br,t),\hat{n}(\br',t')] \vert \Psi(\mT)\rangle$, with  $\hat{n}(\br,t) =  e^{i \hat H^{(0)}t}\hat{n}(\br) e^{-i \hat H^{(0)}t}$, 
which yields
\bea
\nonumber
&\tilde\chi&[n^{(0)}_\mT; \Psi(\mT)](\br,\br',t,t')
% = -i\theta(t-t')\langle \Psi(\mT) \vert [\hat{n}(\br,t),\hat{n}(\br',t')] \vert \Psi(\mT)\rangle
=-i\theta(t-t')\sum_{n,m,k} P_{nm}(\mT)\times\\
&\Big(&f_{nk}(\br)f_{km}(\br')e^{i \frac{(\omega_{nk}+\omega_{mk})}{2}(t-t')}e^{i\omega_{nm}\frac{(t+t')}{2}} -
 (\br \leftrightarrow \br',  t \leftrightarrow t') \Big)
 %(\shortstack{ \br \leftrightarrow \br' \\ t \leftrightarrow t'}) \Big) 
 % shortstack does not help....
%- f_{nk}(\br')f_{km}(\br)e^{i\left(E_k - \frac{E_n+E_m}{2}\right)(t-t')}e^{i(E_n-E_m)\frac{t+t'}{2}} \right)
%(\br \to \br', t \to t')
%&\tilde\chi&[n^{(0)}_\mT, \Psi(\mT)](\br,\br',t,t') = -i\theta(t-t')\langle \Psi(\mT) \vert [\hat{n}(\br),\hat{n}(\br')] \vert \Psi(\mT)\rangle\\
%&=&-i\theta(t-t')\left(\sum_n \sum_k \vert c_n(\mT) \vert^2 f_{nk}(\br)f_{kn}(\br')e^{i(E_n-E_k)(t-t')} + 
%\sum_{n\neq m} \sum_m \sum_k  c_n(\mT) c_m^*(\mT) f_{nk}(\br)f_{km}(\br')e^{i((E_n+E_m)/2-E_k)(t-t')}e^{i(E_n - E_m)(t+t')/2} + other(\br \to \br'...) \right)***
\eea
\label{eq:chi}
where  $f_{jl}(\br) = \langle \Psi_j \vert \hat{n}(\br) \vert \Psi_l \rangle$, $\omega_{jl} = E_j -E_l$,  $P_{jl}(\mT) = c_j^*(\mT)c_l(\mT)$, and $(\br \leftrightarrow \br',  t \leftrightarrow t')$ simply means to exchange $\br$ with $\br'$ and $t$ with $t'$, and vice-versa, in the first term inside the parenthesis.
%The density-operators are  $\hat{n}(\br,t) =  e^{i \hat H^{(0)}t}\hat{n}(\br) e^{-i \hat H^{(0)}t}$. 
A Fourier transform with respect to $\tau = t-t'$ yields
\bea
\nonumber
&\tilde\chi&[n^{(0)}_\mT, \Psi(\mT)](\br,\br',\omega,T) = \sum_n P_{nn}(\mT)\sum_k \frac{f_{nk}(\br)f_{kn}(\br')}{\omega - \omega_{kn}+i0^+}\\
& +&\sum_{k,n\neq m} P_{nm}(\mT)\frac{e^{i\omega_{nm}T} f_{nk}(\br)f_{km}(\br')}{\omega - \frac{\omega_{kn} + \omega_{km}}{2} +i0^+} + c.c.(\omega \to -\omega)
\label{eq:chi(om,T)}
\eea
where $T=\frac{t+t'}{2}$ and $c.c.(\omega \to -\omega)$ denotes the complex conjugate of all terms with $\omega$ replaced by $-\omega$. 

%The non-equilibrium response function $\tilde\chi[n^{(0)}_\mT, \Psi(\mT)](\br,\br',\omega,T)$  describes the response
%depends on the initial state of the system $\Psi(0)$, 
%$\delta n (\br,t)$ of a system in ana rbitrary (generally non-stationary) state $\Psi(\mT)$,  to an external perturbation $\delta v\ext(\br',t')$. 
 %\footnote{Notice that in the standard . linear response formalism theresponse of the system only involves excitations from the ground state.}.  
The poles of $\tilde\chi[n^{(0)}_\mT, \Psi(\mT)](\br,\br',\omega,T)$  have positions
  independent of $\mT$ and are completely determined by the spectrum of
  the unperturbed Hamiltonian. They correspond to excitations and
  de-excitations from the states populated at time $\mT$, $c_n(\mT) \neq
  0$.  Their residues, determining the amplitude and sign of the
  spectral peaks, depend on the state $\Psi(\mT)$ and on transition densities between the eigenstates.  
 The poles in the second term in Eq.~\eqref{eq:chi(om,T)} may look unusual,
 being the average of two energy differences, but these turn into
  simple energy differences once the response function Eq.~\eqref{eq:chidef} is integrated
  against the external potential, and the observable $\delta n
  (\br,t)$ resonates at frequencies of the unperturbed system. Note that when $\Psi(\mT)$ is the ground-state of $H^{(0)}$, Eq.~\eqref{eq:chi(om,T)} reduces to the usual linear response function in Lehmann representation.

%Recent work on the simulation of pump probe and photoemission experiments within real-time TDDFT has revealed that approximate TDDFT, 
%namely the fact that absorption and emission spectra show an unphysical relative shift ~\cite{GBCWR13}. 
%But the physical absorption and emission from a given state has to occur exactly at the same frequency, since its value does not depend on the applied field,
%only on the characteristics of the system.
 %Here, we derive an exact condition that the xc functional must satisfy in
%order to capture the dynamics accurately. We show that adiabatic
%approximations typically violate this condition. 

Turning now to the TDDFT description,  we find a very different picture.
%we show that the form of the non-equilibrium TDDFT response is far less trivial than Eq.~(\ref{eq:chi(om,T)}).  
%The exact TDKS system reproduces the exact
%time-dependent density of the system at all times via single-particle
%orbitals evolving in a one-body Hamiltonian. 
Imagine solving the time-dependent KS equations while the field is on, and let $\Phi(\mT)$ denote the KS state reached at time $t=\mT$ when the field is turned off. 
Unlike the interacting system,
%which, after the field is turned off evolves under the static potential due to the nuclei,$v\ext^{(0)}(\br)$, 
the KS system evolves in a potential
$v\s^{(0)}(\br,t)= v\s[n_\mT^{(0)},\Phi(\mT)](\br,t)$ that typically continues to evolve in time even in the absence of external fields~\cite{MBW02,M12chap,NRL13}.
This is true for the exact KS potential, as well as for approximate
ones,
% In the interacting system, the time-dependence of the density
%$n_\mT^{(0)}$ arises due to the state $\Psi(t=\mT)$ being a
%superposition of eigenstates of the unperturbed interacting
%Hamiltonian, while in the KS system, the time-dependence of
%$n_\mT^{(0)}$ occurs not only from having a superposition state
%$\Phi(\mT)$ but also from time-dependence in the KS potential itself,
 as a consequence of the xc potential being a functional of the time-dependent density. 

%is a functional of the density $n_\mT^{(0)}$ which continues to evolve in
%time when the system is not left in a stationary state after the field is turned off.
%the problem with writing it this way is that the external potential is also a functional of the density, but it is constant, hile the density is not. 

The time-dependence of $v\s^{(0)}(\br,t)$ implies that the eigenvalues of the
instantaneous KS Hamiltonian change in time for $t>\mT$, when either
the exact or approximate functionals are used~\footnote{The fact that the bare resonances of the adiabatic TDKS potential change in time has been
noted before, and referred to in
Ref.~\cite{FHTR11} as dynamical detuning. We emphasize here that this is however
a feature shared by the {\it exact} time-dependent KS potential, not necessarily due to  any
approximation.}. 
But, except for special cases (see  shortly),
these eigenvalue differences are {\it not} the  KS response frequencies, since $H\s^{(0)}= T + v\s^{(0)}(\br,t)$ is time-dependent. 
The non-equilibrium KS response function at
time $t=\mT$,
\ben
\tilde\chi\s[n^{(0)}_\mT,\Phi(\mT)](\br,\br',t,t') = \left.\frac{\delta n(\br,t)}{\delta v\s(\br',t')}\right \vert_{n^{(0)}_\mT,\Phi(\mT)}\;,
\een
has poles in its $(t-t')$-Fourier transform that define the KS response frequencies, and these are typically $\mT$-dependent (for either exact or approximate functionals; see example shortly).
 Because the
  interaction picture here involves a time-dependent
  Hamiltonian, $H\s^{(0)}(t)$,  the
  density-operators involve time-ordered exponentials
%, $\hat{n}(\br,t)  = \hat{T}e^{i\int_0^t d\tau   H\s^{(0}(\tau)}\hat{n}(\br)\hat{T}e^{-i\int_0^t d\tau   H\s^{(0)}(\tau)}$, 
and a simple interpretation of its
  Fourier transform with respect to $(t-t')$,
  $\tilde\chi\s(\br,\br',\omega,T)$, in terms of eigenvalue differences of some  static KS
  Hamiltonian is generally not possible. 
Still, from the fact that the physical and KS systems yield the same density-response, we can derive a Dyson-like equation linking the two 
response functions:
\ben
\tilde\chi^{-1}(\omega,T) = \tilde\chi\s^{-1}(\omega,T) - \tilde{f}\Hxc(\omega,T)
\een
%\bea
%\tilde\chi[n_\mT^{(0)},\Psi(\mT)] &=&\tilde\chi\s[n_\mT^{(0)},\Phi(\mT)]\\
%&+& \tilde\chi\s[n_\mT^{(0)},\Psi(\mT)]\ast\tilde{f}\Hxc[n_\mT^{(0)},\Psi(\mT),\Phi(\mT)]\ast\tilde\chi[n_\mT^{(0)},\Psi(\mT)]
%\eea
 dropping the spatial  arguments and functional dependencies to avoid clutter. 
We defined the generalized Hartree-xc kernel as $\tilde f\Hxc = 1/\vert\br - \br'\vert + \tilde f\xc$, where 
\ben
\tilde{f}\xc[n^{(0)}_\mT;\Psi(\mT),\Phi(\mT)](\br,\br',t,t') = \left. \frac{\delta v\xc(\br,t)}{\delta n(\br',t')} \right\vert_{n^{(0)}_\mT,\Psi(\mT),\Phi(\mT)}\;.
\een
%The response function of Eq.~(\ref{eq:chi(om,T)}) can then be obtained from TDDFT via

%Compared to linear response from a ground-state, there is an additional facet to the task of 
The generalized kernel 
must shift the $\mT$-dependent response frequencies of the KS system to the $\mT$-{\it independent} ones   of the interacting system.
%but must do it in a way that cancels the $\mT$-dependence of the KS frequencies. 
We can now state the exact condition: Let $\omega_i$ be a pole of $\left(\tilde\chi\s^{-1}[n^{(0)}_\mT,\Phi(\mT)] - \tilde f\Hxc[n^{(0)}_\mT,\Psi(\mT),\Phi(\mT)]\right)^{-1}$, then $\omega_i$ should be invariant with respect to $\mathcal{T}$:
\ben 
\frac{d\omega_i}{d\mT} = 0\;.
\label{eq:ex_cond}
\een 
This gives a strict condition that is particularly important in
time-resolved spectroscopic studies~\footnote{For pump-probe spectroscopy, this exact condition states
that the resonant frequencies are independent of the duration of the
pump. It is also true that the frequencies are independent of the
pump-probe delay $\tau$ which can be thought of as a special case of
Eq. 7 where the external field has zero amplitude for
${\mathcal T}-\tau<t<{\mathcal T}$} and in resonant dynamics: in some cases more
important than accuracy in the actual values of the predicted
response frequencies is their invariance with respect to
$\mT$. Approximate kernels may shift the poles of the KS
response function towards the true response frequencies, but unless they cancel the
$\mT$-dependence of the KS poles, they will give erroneously 
$\mT$-dependent spectra.

This has implications even in the cases where the nuclei cannot be
considered as clamped. There, in the physical system, the
electronic excitations couple to ionic motion, so that
the potential $v\ext^{(0)}$, which depends on the nuclear positions,
depends on $\mT$ and on the time delay between pump and probe.
% $\delta v\ext(t')$ is
%applied. This means the energy eigenvalues are functions of these
%times, and 
The time-resolved resonance spectrum can then be interpreted as ``mapping
out'' the potential energy surfaces of the molecule.
 %In the case of pump-probe set-ups,
%But there should be no time-dependent shifts in the positions of the resonances in the time-resolved electronic spectrum when the nuclei do not move. 
Time-dependence should arise purely from ionic motion: spurious
time-dependence in approximate TDDFT simulations arising from
violation of condition~(\ref{eq:ex_cond}) in the limit of clamped
ions will muddle the spectral analysis in the moving-ions case,
and could be mistaken for changes in the nuclear configuration.

The exact satisfaction of condition~(\ref{eq:ex_cond}) is generally
difficult for approximate functionals, but reasonable results could be
obtained if its violation is weak. Shortly, we will give 
examples where a functional approximately satisfies
Eq.~(\ref{eq:ex_cond}) and yields accurate resonant dynamics,
despite an inaccurate value of the resonant frequency. On the other hand, we will find cases where the
response frequency given by an approximate functional is quite
accurate at time $\mT$ but where violation of
condition~(\ref{eq:ex_cond}) leads to a drastic qualitative failure in
the dynamics.

Before turning to examples, consider when the  pumped (interacting) system is in a  stationary excited state, so it has a static density: $n_\mT^{(0)} = n_k$, the density of the excited state. 
Within the adiabatic approximation, the KS potential also becomes constant~\cite{MB01,M12chap}, 
%There are two
%possibilities for the exact KS potential~\cite{MB01,M12,NRL13check}: either (i) it becomes static, but not equal to the ground-state KS potential corresponding
%to $v\ext^{(0)}$, or (ii) it continues to change in  time, with the sum of the time-dependent  orbital-densities  remaining constant. Within an
%  adiabatic approximation, clearly only (i) is possible. Considering then
%  possibility (i),
and we observe that the state $\Phi(\mT)=\Phi_k$ solves
  the self-consistent field (SCF) equations for the static potential
  $v\s^{(0)}[n_k;\Phi_k]$. An expression for $\tilde\chi\s[n_k;\Phi_k](\omega,\mT)$ analogous to Eq.~\eqref{eq:chi(om,T)} can be found, with the poles given by
the eigenvalue differences of the corresponding $H\s^{(0)}$. Denoting these KS frequencies as $\omega_{\sss{S},i}^{k}$ (the $i$th KS frequency as computed from the $k$th SCF state), then the exact condition~\eqref{eq:ex_cond}  can be turned into a condition on a matrix equation directly for the interacting frequencies. 
%\ben
%\Omega_i = \omega_{\sss{S},i}^{k} + corr.\left(\tilde{f}\Hxc\right) \; {\rm independent\;\; of\;} k
%\een
Within a single-pole approximation (SPA), the condition is that
\ben
\omega_i = \omega_{\sss{S},i}^{k} + 2\int dr dr' \phi_i^{k}(\br)\phi_a^{k}(\br)\tilde{f}\Hxc^{k}(\br,\br')\phi_i^{k}(\br')\phi_a^{k}(\br')\;,
\label{eq:spa}
\een
for spin-saturated systems, must be independent of $k$. For spin-polarized systems and non-degenerate KS poles, replace $\tilde f\Hxc$ with $(1/\vert\br - \br'\vert +\tilde f\xc^{\sigma,\sigma})$.
The violation of this condition is responsible for the spurious peak shifting between fluorescence and absorption recently observed in Refs.~\cite{RN12,RN12c,GBCWR13}.

%Note that even in cases where $n_\mT^{(0)}$ is static in the absence
%of external time-dependent fields, the exact
%$v\s[n_\mT^{(0)},\Phi(\mT)](\br,t)$ is typically time-dependent,
%unless $\Phi(\mT)$ is an eigenstate of $v\s(\br,0)$; in this case
%however, the KS potential in an adiabatic approximation is static

We illustrate the consequences of the exact conditions Eqs.~(\ref{eq:ex_cond}) and (\ref{eq:spa}), using the
example of resonant charge-transfer (CT)
dynamics.  A simple model Hamiltonian of two soft-Coulomb interacting electrons in one dimension~\cite{JES88,LEHG99,VIC96,BS02,LGE00,EFRM12,FERM13} allows us to compare with exact results. 
%The unperturbed Hamiltonian has the form 
%\ben
%H = -\half \sum_i^N \frac{d^2}{dx_i^2} + \sum_{i\neq j}^N \frac{1}{\sqrt{1+(x_i-x_j)^2}}  + \sum_i^N v\ext(x_i)
%\label{eq:H}
%\een
We take $v\ext(x) = -2/\sqrt{(x+R/2)^2 + 1} -2.9/\cosh^2(x+R/2) - 1/\cosh^2(x-R/2)$ with $R=7$au and zero boundary conditions at $\pm 50$au. 
%, which in atomic units, are $Z_L = 2,Z_R = 0, U_L =2.9, U_R=1$ and $R = 7$a.u is internuclear distance.

\begin{figure}
	\includegraphics[width=0.5\textwidth, height = 6cm]{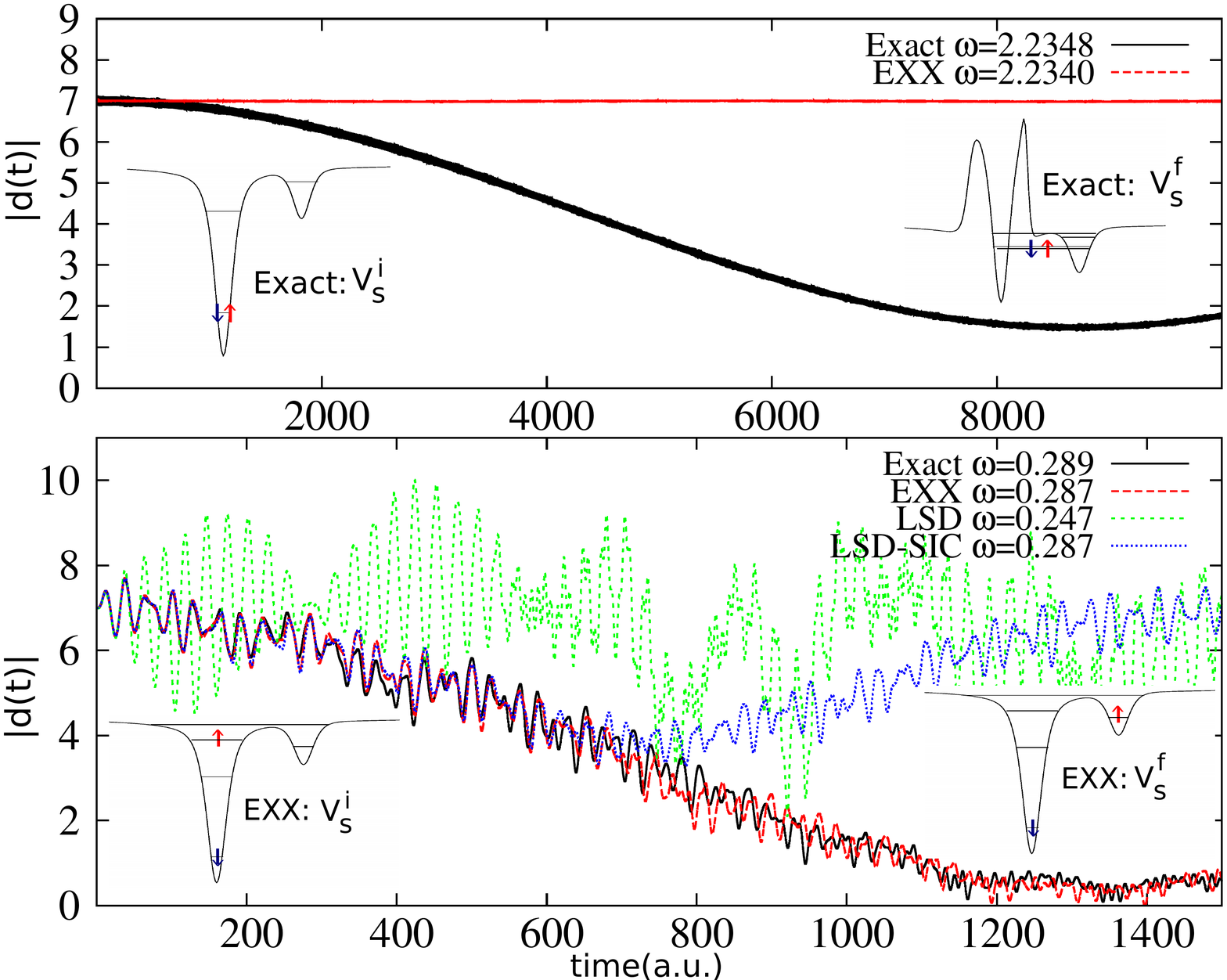}
	\caption{Dipole moments calculated from the center of the double-well: exact (black),  EXX (red), LSD (green) and SIC-LSD (blue), driven at resonant $\omega^i$ for each. The initial and target-final KS potentials are shown as insets, exact in the top panel, and in EXX in the lower panel. Top panel: CT from the ground-state. Lower panel: CT from the ``photo-excited'' state.}
%EXX almost mimics the dynamics of the exact case, as explained in text. However, LSD fails much worse than SIC-LSD, as seen from the large discrepancy between $\omega\s^i$ and $\omega\s^f$.}
	\label{fig:CT}
\end{figure}

Resonant CT beginning in the ground-state provides an example of
dramatically changing KS resonances, even for the exact KS potential.
The ground-state has two electrons in the left well, and the exact
initial KS potential $v\s^i$ is shown on the left in the top panel in Fig.~\ref{fig:CT}. The
KS CT excitation frequency is
$\omega\s^{i} = 2.2348$a.u which happens to equal the true
(interacting) CT excitation, up to the 5th decimal place. If the exact KS
system is driven by a weak-enough resonant field, it achieves  the exact
density of the true CT excited state via a doubly-occupied KS orbital after half a Rabi cycle. The exact KS potential at this final time, $v\s^f$ (on the right of top panel of Fig.~\ref{fig:CT}), looks very different: it displays a step, which,
% of size $\vert I_D^{N-1} - I_A^{N+1}\vert$ 
in the limit of large separation~\cite{FERM13},
results in ``aligning'' the lowest level of each well. Therefore
the KS response frequencies are completely different than those at the
initial time: $\omega\s^{f} = 0.0007$a.u. 
%The final CT state reached
%by propagating with the exact KS potential is a doubly-occupied
%bonding orbital of half the exact density of the interacting CT state.
 $\tilde{f}\Hxc$ plays an increasingly crucial role in maintaining
constant TDDFT response frequencies of Eq.~(\ref{eq:ex_cond}), $\omega^i = \omega^f = 2.2348$au: at first
its effect is small but as the charge transfers, its correction to the
KS response frequency increases dramatically.
The dipole dynamics for field $E(t) = 0.05\sin(2.2348t)$ au is shown. 

Now turning to approximations: the approximate KS resonances also
change in time significantly, but the approximate kernel corrections
are typically small, resulting in grave violations of
condition~(\ref{eq:ex_cond}) and~(\ref{eq:spa}). For example, in exact-exchange (EXX)
$\omega\s^i = 2.2340$ while again $\omega\s^f$ tends to zero, with the
$f\Hx$ correction in the fifth decimal place in both the initial and final states. As a consequence, the EXX dipole dynamics driven at its resonance completely fails to charge transfer, as seen in the top panel of Fig~\ref{fig:CT}.

Other recent works have noted the failure of adiabatic functionals in
TDDFT (including the adiabatically-exact) to transfer charge across a
long-range molecule~\cite{RN11,RP10,FERM13,FM14,FM14b,HRCLG13}, even
when their predictions of the CT energies are very
accurate~\cite{FM14,FM14b}, as computed from the ground-state response. Here we
attribute their failure to the violation of condition~(\ref{eq:spa}), 
as for the case of EXX above. The resonant frequencies
predicted by the functional in the initial state and in the target CT
state are significantly different from each other. This is due to
having one delocalized KS orbital describing the final CT state,
resulting in static correlation in the targeted final KS system, and a
grossly underestimated CT frequency when computed via the response of
the target CT state. The CT frequency computed in the initial
ground-state, on the other hand, can be quite reasonable, as seen above. 
%In some of those works, the poor
%performance was attributed to 
%failure of the build-up of a
%correlation step across the molecule; the latter feature is a consequence of 
%having one doubly-occupied delocalized KS orbital throughout the evolution 
%describing an interacting state that involves minimally two orbitals
%in the target CT state.  

We next consider CT from a singly-excited state 
%One might then ask whether approximations in
%TDDFT do a better job for CT from an excited state, because 
where the
KS system involves more than one orbital, and the transferring
electron is not tied to the same orbital that the non-transferring
electron is in.  Simulations on real systems indeed often start in a photoexcited state\cite{Carlo13}.
% The ground state of our CT model has two electrons of opposite spin
% occupying the left well.  
We consider a ``photoexcitation'' in our model molecule that
 takes the interacting system to its 4th singlet excited state, localized on the left well. We then apply a weak driving field,  $E(t) =0.0067\sin(\omega t)$a.u., at
 frequency $\omega = 0.289$a.u., that is resonant with a CT
 state that has essentially one electron in each well (see lower panel of
 Fig.~\ref{fig:CT}). For this case, $v\s^i$ and $v\s^f$ within EXX are shown; the exact ones are similar. The exact dipole (Fig.~\ref{fig:CT}) shows
 almost complete CT.
%(there is small population transfer  to other states).

We now consider TDDFT simulations of this process, using three
functionals: EXX, local-spin-density approximation
(LSD), and self-interaction corrected LSD (SIC-LSD). For each, we begin the calculation in the 4th excited KS state, as
would be done in practise to model the process above. However, we first relax the state via an SCF
calculation to be a KS eigenstate, so that there is no dynamics until
the field is applied, as in the exact problem. 
%We denote this initial state as $\Phi^*_L$. 
We then apply a weak driving field of the same strength as applied to
the interacting problem, but at the CT frequency of the approximate
functional, computed from the initial state, $\omega^i$. In Table~\ref{tab:CT} one can  contrast this with the
values for the CT frequency computed from the target final CT state,
$\omega^f$, as well as the bare KS eigenvalue differences,
$\omega\s^i$ and $\omega_s^f$.  The approximate TDDFT
corrections to the bare KS values for CT are very small, as expected.
%Although a peak was not discerned in LSD in the dynamic response of the
%initial state, we expect $\omega \approx \omega\s$. 
Most notable is that the CT
TDDFT EXX frequencies, $\omega^i$ and $\omega^f$, computed in the initial and CT states is identical up to the third decimal place, while there is significant difference
amongst the SIC-LSD values, and even more amongst
LSD. In light of the exact conditions~(\ref{eq:ex_cond}) and~(\ref{eq:spa}), we expect EXX
to resonantly CT well, while SIC-LSD would suffer from
spurious detuning, and LSD even more. Indeed, this speculation is
borne out in Fig~\ref{fig:CT} lower panel: EXX captures the exact dynamics
remarkably well. 
SIC-LSD begins to CT but
ultimately fails due to its response frequencies continually changing
during the dynamics, as reflected in the initial and final snapshots
of the frequencies given in the table.  LSD, with its even greater
difference in the initial and targeted-final response frequency, indeed fails miserably.  
%The
%LSD dynamics is mainly due to an excitation into the fifth orbital,
%which has delocalized character, and has a frequency close to the CT
%frequency.  
%Interestingly, if the dynamics is run ``in reverse'',
%i.e. starting in the SCF CT state and driven at the frequency
%$\omega^f$, again EXX shows full CT, SIC-LSD partial, and LSD also
%manages to partially CT (there is no other excitation that interferes
%strongly in this case), but less than the others.  We also tried
%driving the dynamics with different frequencies, e.g. $\omega\s^f$, or
%the $\Delta$SCF value~\cite{ParrYangbook} computed from ground-state
%energy differences but, none of these was successful for LSD or
%SIC-LSD, and reduced the CT for the case of EXX. 
Note that, as in
practical calculations, spin-polarized dynamics is run from the
initial singly-excited KS determinant, with the idea that results
would be spin-adapted at the end.

Why does EXX not suffer from  spuriously time-dependent response
frequencies here? For the special case of two electrons in a
spin-symmetry-broken state, $v^{\rm EXX,\uparrow}\xc =
-v\H[n_\uparrow]$, so $v\s^{\rm EXX,\uparrow} = v\ext +
v\H[n_\downarrow]$. Driving with a weak field resonant with the
$\uparrow$-electron excitation, where the $\uparrow$ is promoted in
the initial state, causes only a gentle jiggling of the
$\downarrow$-electron, so that the $\uparrow$ sees an almost static
potential; in this sense EXX mimics the exact functional, that keeps
the response frequencies static. The bare KS frequency hardly changes (see potentials in lower figure),
and, within the spin-decomposed version of SPA
Eq.~\eqref{eq:spa}, the correction due to the EXX kernel 
vanishes. So, absorption and emission peaks are on top of each other.
For general dynamics, we do not advocate EXX, even for
two-electron systems (see previous example);  it works in this example because of the conditions above that lead to the nearly constant KS potential. 

\begin{center}
\begin{table}
\begin{tabular}{|c|c|c|c|c|c|}\hline
 & $\omega\s^i$ &$\omega\s^f$ & $\omega\s^{\rm g.s.}$ &TDDFT $\omega^i$ &TDDFT $\omega^f$ \\\hline
EXX &  0.286 & 0.286 & 0.288 & 0.287 & 0.287\\ \hline
LSD &0.247 &0.094 & 0.482 & -- & 0.091 \\\hline
SIC-LSD &0.287 & 0.236 & 0.267 & 0.287 & 0.237\\\hline
%TDDFT \omega^i & & 0.287 & no peak & to be done\\\hline
%TDDFT \omega^f & & to be done & to be done & to be done\\\hline
%$\omega ^{ \Delta SCF}_{ CT }$ & & 0.285 & 0.224 & 0.278 \\ \hline
%Eig-diff local exc SCF14 & &  &  & \\\hline
%Eig-diff local exc SCF13 & &  &  & \\\hline
\end{tabular}
\caption{Bare KS and TDDFT-corrected photo-excited CT frequencies computed in the initial, targeted final, and ground states, in atomic units (au). 
%The well parameters, in au, are $Z_L = 2,Z_R = 0, U_L =2.9, U_R=1$ and $R=7$ a.u.  
The exact CT frequency is $\omega  = 0.289$ au. The TDDFT values were obtained via  linear response to a $\delta-$kick perturbation~\cite{YNIB06} and ``--'' indicates no peak was discernible in the spectra, but we expect $\omega^i \approx \omega\s^i$. Calculations were performed using the octopus code~\cite{octopus,octopus2}: a box of size 50 au, grid spacing 0.1au, and time-step 0.005au were used.}
\label{tab:CT}
\end{table}
\end{center}

%Similar observations hold for other dynamics. 
%Revisiting the problem
%of CT when starting in the ground-state in
%Refs.~\cite{FERM13,FM14,FM14b},  we can attribute the failure of
%approximate TDDFT functionals to the violation of
%condition~(\ref{eq:spa}): The CT resonant frequencies calculated via
%response of the initial ground-state by EXX, LSD, SIC-LSD, and
%adiabatically-exact, vary greatly from that predicted by the
%corresponding functional evaluated via response of the final
%target-state.  This is related to having one delocalized KS orbital
%describing the final CT state, resulting in static correlation in the
%targeted final KS system, and a grossly underestimated CT frequency
%when computed via the response of the target CT state. The CT
%frequency computed in the initial ground-state, on the other hand, can
%be quite reasonable. 

%build-up of a step feature due to having only one
%occupied KS orbital. The step induces static correlation in the target
%CT state, resulting in approximate TDDFT functionals grossly
%underestimating the CT frequency in the target CT state, while doing a
%reasonable job in the ground-state.

In a third example, when
resonantly driving between two locally excited states in a single
well ($v\ext = -2/\sqrt(x^2+1) - 3/\cosh^2x$), one finds again  that the EXX frequencies computed from each
excited state are very similar, $0.824$au,  quite different than the exact resonant frequency of $0.755$au. Despite this large discrepancy, the EXX dipole closely follows the exact one, due to the approximate satisfaction of condition~(\ref{eq:spa}), and, likely, condition~(\ref{eq:ex_cond}),
%Fig.~\ref{fig:local} shows resonant dynamics to a local excitation in
%a single well (see inset)~\footnote{KS populations usually only have very limited meaning given that the KS wavefunction is not supposed to mimic the exact one. However in this case of targetting two-level dynamics, starting with two-orbitals, they are useful to analyse the dynamics.}.
%: take $U_{L} =0,Z_L=0,U_R=3, Z_{R}=2,R=0$ in the external potential. 
%Table~\ref{tab:local} gives their energies, computed analogously to
%the CT case. Again, the EXX frequencies remain nearly constant through
%the evolution, as suggested by the similarity of the bare KS frequencies in the initial and final states, but our numerical resolution was not fine enough to obtain the TDDFT $\omega^i$ to the third decimal place in this case. EXX predicts
%accurate dynamics, as expected from condition~(\ref{eq:ex_cond}) (or Eq.~(\ref{eq:spa}))despite the inaccurate resonant frequency compared with the exact.
%SIC-LSD  again yields a poorer dipole dynamics  due to its spuriously time-dependent resonance, as suggested by the larger difference in $\omega^i$ and $\omega^f$.  
LSD again violates condition~(\ref{eq:spa})  the most severely, and its dynamics is consequently the worst.

Interestingly, our exact condition could explain the success of the
``instantaneous ground-state'' approximation over the adiabatic
approximation, explored in Ref.~\cite{MRHG14}: there, for initial
non-stationary states evolving in a time-independent external field,
the KS potential is always taken as equal to the initial one, so has
static resonances, satisfying Eq. (7).

%\begin{center}
%\begin{table}
%\begin{tabular}{|c|c|c|c|c|c|}\hline
% & $\omega\s^i$ &$\omega\s^f$ & $\omega\s^{\rm g.s.}$ &TDDFT $\omega^i$ &TDDFT %$\omega^f$ \\\hline
%EXX &  0.824 & 0.824 & 0.825 & 0.82 & 0.829\\ \hline
%LSD &0.857 &0.738 & 0.807 & 0.856 & 0.74 \\\hline
%SIC-LSD &0.826 & 0.860 & 0.821 & 0.81  & 0.837 \\\hline
%\end{tabular}
%\caption{Bare KS and TDDFT-corrected local-excitation frequencies for the three functionals computed in the initial, targetted final, and ground states, in atomic units (au). The well parameters, in au, are $Z_L = 0,Z_R = 2, U_L =0, U_R=3$ and $R=0$ a.u. The exact frequency is $\omega  = 0.755$ au. }
%\label{tab:local}
%\end{table}
%\end{center}

%\begin{figure}
% height = 5 cm looks a bit squeezed, 7cm better?
%   \includegraphics[width=0.5\textwidth, height = 7cm]{local_down.pdf}
%   \caption{Dipoles and populations of exact and TDDFT results for the local dynamics where parameters are $U_L = 0, Z_L = 0,U_R = 3,Z_R = 2,R = 0$. {\color{blue}need clarification of chosen freqs...}}
   %Give exact dipole (black), EXX, LSD, SIC-LSD each at the $\Omega^i$. Inset with cartoon of double-well and e transferring. The population of the target excited state is also shown.}
%       \label{fig:local}
%\end{figure}

In conclusion, we have derived a new exact condition that should be
satisfied by approximate functionals in TDDFT in order to accurately
capture non-equilibrium dynamics.  Violations of this condition lead
to misleading results in simulating time-resolved spectroscopy, and failure in
resonantly driven processes. We have shown that even if a functional
does not yield accurate excitation frequencies, if these frequencies
even approximately satisfy the exact condition Eq.~(\ref{eq:ex_cond}) then the predicted
non-linear dynamics could still be accurate.  
%We found
%that the condition is approximately satisfied by adiabatic spin-polarized EXX
% for some 2-electron cases; whether it can be satisfied within an
%adiabatic approximation in a general case remains to be explored.  
The
effect of the spurious time-dependent resonances of approximate
functionals for realistic systems could  be dampened, due
to the large number of electrons and vibronic couplings, but further
investigations are necessary. Likely for spectroscopy or resonant
control processes, satisfaction of the exact condition is essential, and our findings explain related observations in the real systems studied in Refs.~\cite{RN12,RN12c,HTPI14,GBCWR13}.
 The
exact condition highlights a new feature that must be considered in
the development of improved functionals to be able to accurately
capture dynamics far from the ground-state.

%Therefore 
%{\bf keeping the resonances of the electronic system constant is an exact condition that the frequency-dependent xc functional must satisfy.}
%We show that approximate adiabatic xc functionals  will typically predict resonances that change in time, with dramatic consequences in the 
%simulated dynamics. The unphysical shift between the peaks of absorption and emission spectra reported in 
%Ref.~\cite{GBCWR13} can be claimed to this inexactness.

\begin{acknowledgments} We thank Hardy Gross for  lively and helpful discussions.  Financial support from the National Science Foundation
CHE-1152784 (for K.L.), Department of Energy, Office of Basic Energy
Sciences, Division of Chemical Sciences, Geosciences and Biosciences
under Award DE-SC0008623 (N.T.M., JIF), a grant of computer time from
the CUNY High Performance Computing Center under NSF Grants
CNS-0855217 and CNS-0958379, and the RISE program at Hunter College,
Grant GM060665 (ES), are gratefully acknowledged.
\end{acknowledgments}

\bibliography{ref}{}
\bibliographystyle{apsrev4-1}

\end{document}